\documentclass[aps,prl,twocolumn,superscriptaddress]{revtex4-2}
\usepackage{graphicx}
\usepackage{epstopdf}
\usepackage{bm}
\usepackage{latexsym} 
\usepackage{braket}
\usepackage[colorlinks,linkcolor=blue,citecolor=blue,urlcolor=blue]{hyperref}
\usepackage{amssymb}
\usepackage{amsmath}
\usepackage{mathtools}
\usepackage{color}
\usepackage{import}
\usepackage[caption = false]{subfig}
\usepackage{import, xcolor}
\usepackage[retainorgcmds]{IEEEtrantools}
\usepackage{bm}
\usepackage[none]{hyphenat}
\usepackage{textcomp} 

\newcommand{\D}{\mathrm{d}}
\newcommand{\beq}{\begin{equation}}
\newcommand{\eeq}{\end{equation}}

\newcommand{\un}[1]{\mathrm{\:#1}} 
\renewcommand{\vec}[1]{\boldsymbol{#1}} 
\newcommand{\ii}{i}

\begin{document}

\title{Anomalous Optical Drag}
\author{Chitram Banerjee}
\affiliation{Department of Physics of Complex Systems, Weizmann Institute of Science, Rehovot 76100, Israel}
\author{Yakov Solomons}
\affiliation{Department of Physics of Complex Systems, Weizmann Institute of Science, Rehovot 76100, Israel}

\author{A. Nicholas Black}
\affiliation{Department of Physics and Astronomy, University of Rochester, Rochester, NY 14627, USA}

\author{Giulia Marcucci}
\affiliation{Department of Physics, University of Ottawa, Ottawa, ON K1N 6N5, Canada}

\author{David Eger}
\affiliation{Department of Physics of Complex Systems, Weizmann Institute of Science, Rehovot 76100, Israel}

\author{Nir Davidson}
\affiliation{Department of Physics of Complex Systems, Weizmann Institute of Science, Rehovot 76100, Israel}

\author{Ofer Firstenberg}
\affiliation{Department of Physics of Complex Systems, Weizmann Institute of Science, Rehovot 76100, Israel}

\author{Robert W. Boyd}
\affiliation{Institute of Optics, University of Rochester, Rochester, NY 14627, USA}
\affiliation{Department of Physics, University of Ottawa, Ottawa, ON K1N 6N5, Canada}

\begin{abstract}
A moving dielectric medium can displace the optical path of light passing through it, a phenomenon known as the Fresnel-Fizeau optical drag effect. The resulting displacement is proportional to the medium's velocity. In this article, we report on an anomalous optical drag effect, where the displacement is still proportional to the medium's speed but along the direction opposite to the medium's movement. We conduct an optical drag experiment under conditions of electromagnetically-induced transparency and observe the transition from normal, to null, to anomalous optical drag by modification of the two-photon detuning.
\end{abstract}
\maketitle
\section{Introduction}
As light travels through a moving medium, it is dragged along the axis of the medium's motion.  This effect, known as optical drag, was first described theoretically by Augustin-Jean Fresnel in 1818~\cite{1818Fresnel, 1972Schaffner}. In 1851, Hippolyte Fizeau demonstrated Fresnel's drag experimentally~\cite{1851Fizeau}.  However, Fresnel and Fizeau ignored the effect of refractive index dispersion. This effect was incorporated by Hendrik Lorentz, who in 1904 predicted the influence of dispersion on the optical drag effect for a moving medium with a fixed boundary~\cite{1904Lorentz}.  Four years later, Jakob Laub developed a theoretical treatment for optical drag in a dispersive moving medium with a moving boundary~\cite{1908Laub}.
Many experiments that measured the effect of dispersion on optical drag were then reported by Zeeman and coauthors~\cite{1914Zeeman,1915Zeeman,1919Zeeman,1920Snethlage,1922Zeeman}.
They measured the wavelength dependence of optical drag in water~\cite{1914Zeeman,1915Zeeman} and performed other experiments to measure optical drag in quartz and flint glass~\cite{1919Zeeman,1920Snethlage,1922Zeeman}.

The transverse displacement of a light beam experiencing optical drag depends on the both the refractive and group index of the beam in the moving medium~\cite{2003Carusotto}.  It is well known that materials can be highly dispersive under nearly-resonant excitation~\cite{2020Boyd}.  This large dispersion leads to a large group index, to highly subluminal pulse propagation (sometimes referred to as slow light), and consequently to an enhancement of the optical drag effect~\cite{2011Franke-Arnold,2016Safari}.  Since absorption is typically high under nearly-resonant excitation, most observations of slow light have relied upon electromagnetically-induced transparency~(EIT)~\cite{2008Firstenberg,2009Firstenberg} or coherent population oscillation~\cite{2003BigelowSC,2003BigelowPRL,2007Piredda}.  These effects have the added benefit of providing very narrow resonances, leading to high dispersion and large group indices.  As a result, EIT has been used to further enhance the optical drag effect \cite{2020Solomons} and perform high precision velocimetry \cite{2020Chen}.

Under conditions of anomalous dispersion, the group index can become negative, leading to the so-called ``fast light'' effect, where the peak of a pulse advances as it travels through a dispersive medium instead of accumulating a delay~\cite{2009Boyd}.  Under certain circumstances, the advancement of the peak can be attributed to pulse reshaping caused by saturated absorption or gain~\cite{1966BasovD,1966BasovJ}. 

An experimental demonstration of optical drag under conditions of anomalous dispersion has not been previously reported. In this article, we report on the anomalous optical drag effect, where the light beam shifts in the direction opposite to the motion of the medium. We excite EIT in a moving cell of rubidium vapor and induce a transition from normal to anomalous optical drag by properly modifying the two-photon detuning.
\section{Theoretical Model}
\begin{figure}
\centering
\includegraphics[width=\columnwidth]{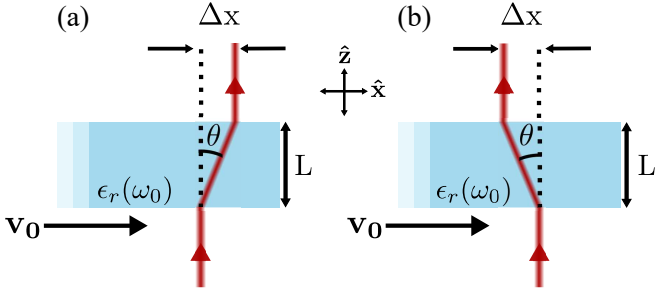}
\caption{Pictorial representation of the transverse Fresnel-Fizeau drag effect for (a) normal optical drag, and (b) anomalous optical drag.}
\label{fig:drag}
\end{figure}

The transverse Fresnel-Fizeau drag effect in a generic dispersive medium has been treated theoretically by starting from the linearized Lorentz transformation for the frequency and wave vector of a beam propagating inside a moving medium~\cite{2003Carusotto}.
Let us consider a nonmagnetic and isotropic medium (i.e., relative magnetic permeability $\mu=1$ and dielectric tensor $\epsilon_{ij}=\epsilon\delta_{ij}$), with $\epsilon_r=\Re[\epsilon]$. 
As shown in Fig.~\ref{fig:drag}(a), if the medium is moving with velocity $\vec{\mathrm{v_0}}=\mathrm{v_{0}\hat{\vec{x}}}$, where $\mathrm{z}$ is the longitudinal and $\mathrm{x,y}$ are the transverse coordinates, the probe light beam experiences a shift along $\mathrm{x}$ of
\beq
\mathrm{\Delta x}=\mathrm{L}\tan\theta.
\label{eq:Delta}
\eeq
In Eq.~(\ref{eq:Delta}), $\mathrm{L}$ is the medium's longitudinal length, and $\theta$ the light's walk-off angle inside the medium, with 
\beq
\tan\theta=\frac{\mathrm{v_{0}}}{\mathrm{c}}\left(\frac{\mathrm{c}}{\mathrm{v_g}}-\frac{\mathrm{v_p}}{\mathrm{c}}\right),
\label{eq:dragangle}
\eeq
where $\mathrm{c}$ is the speed of light in vacuum~\cite{2003Carusotto}.  The group velocity and phase velocity of light in the medium are given by
\beq
\mathrm{v_g}=\frac{\mathrm{\mathrm{c}}}{\sqrt{\epsilon_\mathrm{r}(\omega_0)}+\frac{\omega_0}{2\sqrt{\epsilon_\mathrm{r}(\omega_0)}}\left(\frac{\D\epsilon_\mathrm{r}}{\D\omega}\right)_{\omega_0}},
\label{eq:groupv}
\eeq
\beq
\mathrm{v_p}=\frac{\mathrm{c}}{\sqrt{\epsilon_\mathrm{r}(\omega_0)}},
\label{eq:phasev}
\eeq
respectively.

Equation~(\ref{eq:dragangle}) can be expressed in terms of the medium's refractive index $\mathrm{n}(\omega)=\sqrt{\epsilon_\mathrm{r}(\omega)}$, or after fixing the value of the carrier frequency, as $\mathrm{n_0}=\sqrt{\epsilon_\mathrm{r}(\omega_0)}$. Equations~(\ref{eq:groupv},\ref{eq:phasev}) can then be rewritten as $\mathrm{v_g}=\mathrm{c/{n_{g,0}}}$ and $\mathrm{v_\mathrm{p}= c/n_0}$, where $\mathrm{n_{g,0}=n_0+\left(\omega\frac{\D n}{\D\omega}\right)_{\omega_0}}$ is the group refractive index.
The resulting expression, 
\beq
\mathrm{\tan\theta=\frac {v_{0}}{c}\left(n_{g,0}-\frac{1}{n_0}\right)}, 
\eeq
reveals the existence of three different regimes:\\
(i) $\mathrm{n_{g,0}>n_0^{-1}}$ normal (or positive) optical drag [Fig.~\ref{fig:drag}(a)];\\
(ii) $\mathrm{n_{g,0}=n_0^{-1}}$ absence of optical drag;\\
(iii) $\mathrm{n_{g,0}<n_0^{-1}}$ anomalous (or negative) optical drag [Fig.~\ref{fig:drag}(b)]. For a beam experiencing normal dispersion, if $\mathrm{v_{p} \approx c}$ and $\mathrm{v_{g} \ll c}$, the drag can be approximated in terms of the group delay, $\tau$: $\mathrm{\Delta x \approx Lv_{0}/v_{\mathrm{g}} = \tau v_{0}}$ \cite{2020Solomons}.  The conditions of anomalous dispersion, $\mathrm{n_{g,0} < 0}$, satisfy regime (iii), hence the name \textit{anomalous} drag.  In highly dispersive conditions, the relation $\mathrm{\Delta x \approx \tau v_{0}}$ is still valid because $\mathrm{n_{g,0}} \ll 0$. 

In EIT, the transition from normal to anomalous drag occurs close to the transition from negative to positive two-photon detuning,
$\delta$. To model optical drag under EIT conditions, we begin from Eq.~(50) in \cite{2008Firstenberg}, which describes the steady-state propagation of a probe beam under EIT conditions in a gaseous medium of diffusing atoms.  The medium's velocity, $\vec{\mathrm{v}}_0$, is accounted for by adding a nonzero mean to the Boltzmann distribution of atomic velocities with thermal velocity $\mathrm{v_{\mathrm{th}}=(k_\mathrm{B} T/m)^{1/2}}$,
\beq
F(\vec{\mathrm{v}})=\frac{\exp\left(-\frac{(\vec{\mathrm{v}}-\vec{\mathrm{v_0}})^2}{\mathrm{2v_{\mathrm{th}}^2}}\right)}{\mathrm{2\pi v_{\mathrm{th}}^2}}.
\label{eq:boltzmann}
\eeq
$\mathrm{k_\mathrm{B}}$ is the Boltzmann constant, $\mathrm{T}$ is the temperature, and $\mathrm{m}$ is the atomic mass.
Assuming negligible diffraction, the dynamics of the probe beam are described by
\begin{widetext}
\beq
\mathrm{\left[-\ii\delta+\gamma+K\left|\tilde{\Omega}_2(\vec{\mathrm{r}})\right|^2-D\left(\partial_{\vec{\mathrm{r}}}-\ii\vec{\mathrm{\Delta q}}\right)^2+\left(\partial_{\vec{\mathrm{r}}}-\ii\vec{\mathrm{\Delta q}}\right)\cdot\vec{\mathrm{v_0}}\right]\left[\alpha\tilde{\Omega}_2(\vec{\mathrm{r}})\right]^{-1}\left(\partial_z+\alpha\right)\tilde{\Omega}_1(\vec{\mathrm{r}})=K\tilde{\Omega}_2^*(\vec{\mathrm{r}})\tilde{\Omega}_1(\vec{\mathrm{r}})}.
\label{eq:rabi_real}
\eeq
\end{widetext}
Here, $\tilde{\Omega}_{1,2}$ represent the Rabi frequency envelopes in the slowly varying envelope approximation for the probe and control beams, respectively, and $\mathrm{K}$ is the one-photon complex spectrum with total linewidth $\Gamma$ incorporating homogeneous dephasing and Doppler broadening of the optical transition. Moreover, $\gamma$ is the decoherence rate of the ground-state transition, $\mathrm{D}$ is the spatial diffusion coefficient, $\vec{\mathrm{\Delta q}}=\vec{\mathrm{q}}_2-\vec{\mathrm{q}}_1$ is the probe-control wave vector mismatch (with absolute value $\mathrm{\Delta q}$ along $\hat{\vec{\mathrm{x}}}$), and $\alpha$ is the attenuation (per unit distance) without EIT.

\begin{figure*}[!ht]
    \centering
    \includegraphics[width = 0.8\textwidth]{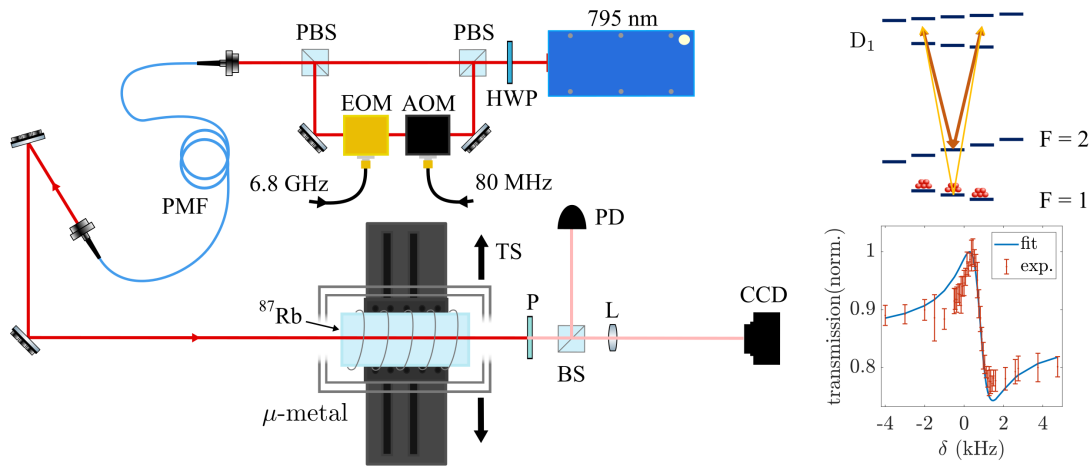}
    \caption{Experimental setup for observing anomalous optical drag under conditions of EIT.  (left) The probe and control fields are derived from a diode laser ($795 \un{nm}$) tuned to the $\mathrm{D}_1$ transition in $^{87}\mathrm{Rb}$.  A half-wave plate (HWP) controls the relative power of the orthogonally polarized probe and control fields before they are split by a polarizing beamsplitter.  The probe passes through an electro-optic modulator (EOM) and an acousto-optic modulator (AOM) to tune it to the $\mathrm{F} = 1 \rightarrow \mathrm{F'} = 1,2$ transition and vary the two-photon detuning, $\delta$, respectively.  The probe and control are then recombined at a PBS before being coupled into a single-mode polarization-maintaining fiber (PMF).  After exiting the PMF both fields pass through a $^{87}\mathrm{Rb}$ cell that is shielded from stray magnetic fields by a double-layer magnetic shield ($\mu\text{-metal}$).  A solenoid creates a uniform magnetic field to lift the degeneracy of the Zeeman sublevels.  The probe and control fields interact through an EIT scheme shown in the upper right panel.  The dominant interaction occurs around the $\mathrm{m_F} = 0 \longrightarrow \mathrm{m^{\prime}_{F}} = \pm 1$ transitions for the probe (yellow arrows) and control (brown arrows).  By detuning both the probe and control $-250 \un{MHz}$ below the $\mathrm{F}^{\prime} = 2$ excited state, the two-photon lineshape is sufficiently asymmetric to produce anomalous dispersion (lower right panel).  The normalized transmission data were fit simultaneously to all other data sets using $\Gamma$, $\gamma$, and $\mathrm{\Delta q}$ as free parameters.  The entire Rb cell apparatus is mounted on a translation stage (TS) that travels transverse to the beam propagation.  A polarizer (P) after the cell filters out the control field, and the probe field is split into two paths by a beamsplitter (BS). One path terminates on a photodiode (PD) for measuring the group delay of the probe.  In the other, a lens (L) images the output facet of the cell onto a camera (CCD).}
    \label{fig:expsetup}
\end{figure*}

In Fourier space, stressing that the medium velocity is parallel to the $\mathrm{x}$-axis, one obtains
\beq
\mathrm{\partial_z\Omega_1=\ii\frac{k_z}{2}\chi(k_x,k_z,\delta)\Omega_1}
\label{eq:rabi_FFT}
\eeq
with the susceptibility derived from Eq.~(\ref{eq:rabi_real}),
\begin{widetext}
\beq
\mathrm{\chi(k_x,k_z,\delta)=\ii\frac{2\alpha}{k_z}\left[1-\frac{K|\Omega_2|^2}
{-\ii\delta+\gamma+K\left|\Omega_2\right|^2+D\left(k_x+\Delta q\right)^2-\ii\left(k_x+\Delta q\right)v_{0}}\right]}.
\label{eq:susceptibility}
\eeq
\end{widetext}
The probe susceptibility's dependence on $\mathrm{k_z}$ is such that $\mathrm{\partial_{k_z}\left[k_z\chi(k_x,k_z,\delta)\right]=0}$.  It should be noted that Eq.~(\ref{eq:rabi_FFT}) is exact only when the control field is an infinite plane wave, $\Omega_2(\vec{\mathrm{r}}) = \Omega_{2}$.  Nonetheless, quantitative agreement between Eq.~(\ref{eq:rabi_FFT}) and experiment can be achieved using a control field of finite extent as long as its beam waist, $\mathrm{w_0}$, is wide enough to satisfy $\mathrm{w_0} \gg \mathrm{max\{\sqrt{D\tau}, v_{0}\tau\}}$~\cite{2020Solomons}.

The resulting solution of Eq.~(\ref{eq:rabi_real}) is
\beq
\mathrm{\tilde{\Omega}_1(x,z)}=\mathcal{F}^{-1}\left\{\mathcal{F}\left[\mathrm{\tilde{\Omega}_1|_{z=0}}\right]\exp\left[\ii\mathrm{\frac{\chi(k_x,k_z,\delta)}{2}k_z z}\right]\right\},
\label{eq:rabi_probe}
\eeq
where $\mathcal{F}$ is the Fourier transform operator.

This picture furnishes an expression for the transverse shift in Eq.~(\ref{eq:Delta}), where the dependence of the optical drag and group delay on the two-photon detuning appears explicitly~\cite{2020Solomons}:
\beq
\mathrm{\Delta x=\partial_{k_x}\Re\left(\frac{\chi(k_x,k_z,\delta)}{2}k_z L\right)},
\label{eq:DeltaEIT}
\eeq
\beq
\mathrm{\tau=\partial_{\delta}\Re\left(\frac{\chi(k_x,k_z,\delta)}{2}k_z L\right)}.
\label{eq:delayEIT}
\eeq
\section{Experimental Results}
The experimental arrangement for measuring anomalous optical drag is shown in Fig.~\ref{fig:expsetup}.  The control and the probe beams are obtained from the same diode laser whose central frequency is set to $795\un{nm}$ to excite the D1 transition of the $^{87}$Rb atoms. The probe is tuned to the $\mathrm{F = 1 \rightarrow F' = 1,2}$ transition, and the control is tuned to the $\mathrm{F = 2 \rightarrow F' = 1,2}$ transition.  Both fields are detuned by $-250 \un{MHz}$ below the $\mathrm{F'=2}$ level, which ensures a sufficiently asymmetric transmission lineshape around the two-photon resonance (see Fig.~\ref{fig:expsetup}, lower right panel), resulting in negative group velocity~\cite{2004Mikhailov}.  A normalized version of Eq.~(\ref{eq:rabi_probe}) was fit to the transmission data simultaneously to all other data sets reported here.
\begin{figure}[!t]
    \centering
    \includegraphics[width = \columnwidth]{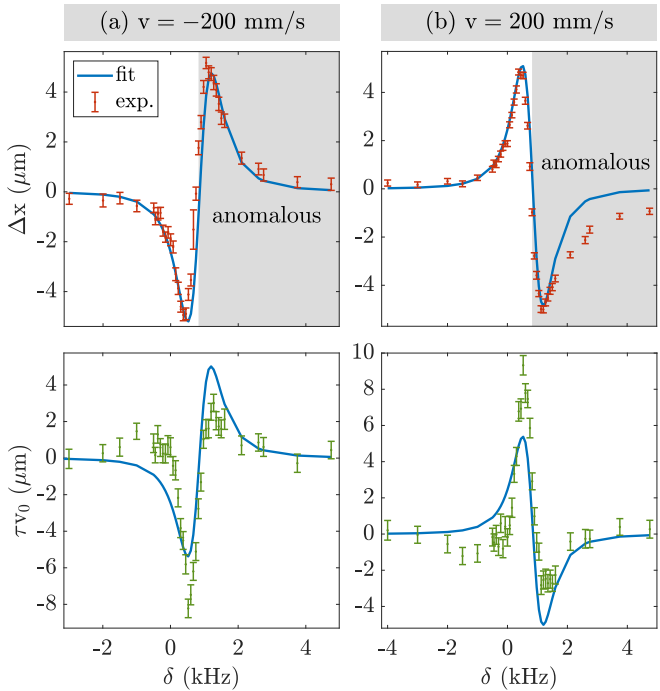}
    \caption{Dependence of optical drag on two-photon detuning $\delta$. (top row) The measured transverse beam displacement $\mathrm{\Delta x}$ is in the same direction as the velocity of the cell on the  red-detuned side of the two-photon transition and in the opposite direction on the blue-detuned side.  This behavior is observed for both (a) negative and (b) positive cell velocities. (bottom row)  The drag expected from the measured group delay, $\mathrm{\Delta x \approx \tau v_{0}}$ is measured simultaneously with $\mathrm{\Delta x}$ and predicts substantially less anomalous optical drag.  The fit lines are obtained from maximum likelihood estimation using the models of Eqs.~(\ref{eq:DeltaEIT},~\ref{eq:delayEIT}) with $\Gamma$, $\gamma$, and $\mathrm{\Delta q}$ as free parameters. An offset of $425 \un{Hz}$ is added to the fit lines to account for a systematic error in the detuning measurement.}
    \label{fig:dragvsdetune}
\end{figure}
The probe and control beams exit the same polarization-maintaining fiber with a radius $\mathrm{w_p = w_c = 2 \un{mm}}$. The total control power is $\sim 40 \un{\mu W}$ and that of the probe is $\sim 2 \un{\mu W}$. The $^{87}$Rb vapor is kept in a magnetically-shielded Pyrex cell of length $\mathrm{L} = 7.5 \un{cm}$. The cell contains buffer gases $\mathrm{N_2}$ ($10\un{Torr}$) and Ar ($90\un{Torr}$) and is heated to $55\un{^{\circ}C}$, in which case the thermal motion of the atoms behaves diffusively with a diffusion coefficient of $\mathrm{D} \approx 166 \un{mm^2/s}$. The cell assembly is mounted on a Thorlabs DDSM100 motorized stage, so it can move with a constant velocity while crossing the beams. 

A polarizer is placed after the cell to filter the probe from the control field.  The probe field is then split into two paths, one that forms an image of the probe at the output of the cell on a Princeton ProEM camera and another that terminates at a photodetector for temporal detection of the probe pulse.  The imaging path is used to measure $\mathrm{\Delta x}$, and the other is used to measure $\tau$.  By comparing the transverse beam profile's center of mass when the cell is moving and not moving, the transverse drag is obtained.  Similarly, the mean temporal shift between pulses for various two-photon detunings gives the group delay of the pulse inside the cell.

The dependence of optical drag on the two-photon detuning is shown in Fig.~\ref{fig:dragvsdetune}.  Both $\mathrm{\Delta x}$ and $\mathrm{\tau v_{0}}$ are measured at cell velocities of $200 \un{mm/s}$ (a) and $-200 \un{mm/s}$ (b), the maximum achievable velocities of our system.  The models of Eqs.~(\ref{eq:DeltaEIT}, \ref{eq:delayEIT}) are fit to the $\mathrm{\Delta x}$ and $\mathrm{\tau v_{0}}$ data respectively using maximum likelihood estimation (MLE) with three free parameters: $\Gamma$, $\gamma$, and $\mathrm{\Delta q}$. Furthermore, we assume that the data have Gaussian random errors.  In fact, all data sets shown were fit simultaneously using MLE, obtaining the following values of the three fit parameters: $\Gamma = 266(2) \un{MHz}$, $\gamma = 145(2) \un{Hz}$, and $\mathrm{\Delta q} = 1.2(3)*10^{-6} \un{2\pi/\lambda}$. Uncertainties in the fit parameter values were obtained through Monte Carlo simulation.  The data show a strong dependence on the two-photon detuning $\delta$; normal optical drag occurs for $\delta < 425\un{Hz}$, while anomalous optical drag occurs for $\delta > 425\un{Hz}$ and peaks around $\delta = 900\un{Hz}$.  Note that the $\mathrm{\tau v_{0}}$ data predict approximately half as much anomalous optical drag than is actually measured, $\mathrm{\Delta x}$.  The breakdown of this approximation may result from systematic error in the estimation of the pulse delay.

\begin{figure}[!t]
    \centering
    \includegraphics[width = \columnwidth]{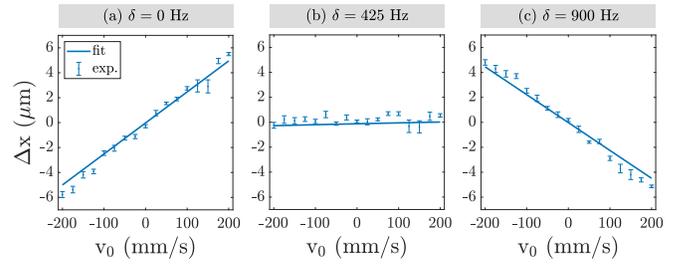}
    \caption{Experimental observation of optical drag at various dragging velocities.  (a)  At a two-photon detuning of $\delta = 0 \un{Hz}$ the beam experiences normal optical drag.  The dependence of the drag on the velocity is approximately linear from $\mathrm{v_{0}} = -200 \un{mm/s}$ to $200 \un{mm/s}$. (b) There is effectively no optical drag for any cell velocity when $\delta = 425 \un{Hz}$. (c) At $\delta = 900 \un{Hz}$ the beam undergoes anomalous optical drag for all velocities in the domain $\mathrm{v_{0}} = -200 \un{mm/s}$ to $200 \un{mm/s}$. The fit lines were obtained by fitting Eq.~(\ref{eq:DeltaEIT}) to all data sets simultaneously.}
    \label{fig:dragvsvelocity}
\end{figure}

Figure~\ref{fig:dragvsvelocity} shows the dependence of optical drag on dragging velocity.  Despite the nonlinear dependence of Eq.~(\ref{eq:susceptibility}), and consequently Eq.~(\ref{eq:DeltaEIT}), on the dragging velocity, the data are approximately linear in this experimental regime.  Interestingly, for a two-photon detuning of $\delta = 425 \un{Hz}$, almost no optical drag is present for any dragging velocity, as shown in Fig.~\ref{fig:dragvsvelocity}(b).  The null optical drag is unique and corresponds to conditions of zero group delay in the medium (zero crossing in Fig.~\ref{fig:dragvsdetune}).  Normal [Fig.~\ref{fig:dragvsvelocity}(a)] and anomalous [Fig.~\ref{fig:dragvsvelocity}(c)] drag are observed over a range of velocities with a high degree of symmetry between the normal and anomalous effects.
\section{Conclusions}
We demonstrated normal, anomalous, and null optical drag effects experimentally and modeled them theoretically.  By using electromagnetically-induced transparency in $^{87} \mathrm{Rb}$ vapor we were able to maximize the dispersion and minimize the loss associated with near-resonant atomic excitation. Our measurements showed that the beam's displacement is proportional to the velocity of the medium through which it passes, with a direction depending on the sign of medium's dispersion.  For positive group delays, the beam experiences normal optical drag and is dragged in the direction of the material motion.  When the measured group delay is negative, the beam is dragged in a direction opposite to the material displacement. In a regime where the group index is exactly equal to the inverse of the refractive index and there is no measured group delay, there is no optical drag at any material velocity.  To our knowledge this is the first experimental demonstration of anomalous drag.

In the gaseous system we study, the average linear velocity of the atoms due to the transverse motion of the cell adds to their underlying thermal velocity distribution, resulting in a drift-diffusive atomic motion. The atomic diffusion can lead to paraxial diffusion of the light field~\cite{FirstenbergRMP13, FirstenbergPRL10} and also, similarly to anomalous and null drag, to effective paraxial diffraction that eliminates and even reverses the free-space optical diffraction~\cite{FirstenbergNatPhys09}.  While here we attenuate the effect of diffusion by using a dense buffer gas, it could be interesting to explore the interplay between (linear) drag and (quadratic) diffusion and diffraction in the anomalous and null regimes.
\begin{acknowledgements}
The authors would like to acknowledge M. Zahirul Alam for very enlightening discussions during the process of conducting this experiment and writing the manuscript.  This work was supported by the Pazy Foundation, the Israel Science Foundation, and the Consortium for quantum sensing of Israel Innovation Authority.
\end{acknowledgements}
C.B. and Y.S. contributed equally to this work.
\end{document}